\documentclass[onecolumn,preprintnumbers,amsmath,amssymb]{revtex4}
\usepackage{graphicx}

\newcommand{\be}{\begin{equation}}
\newcommand{\ee}{\end{equation}}
\begin{document}
\thispagestyle{empty}

\title{Energy loss at zero temperature from extremal black holes}
 \author{\large Moslem Ahmadvand}
 \email{email:ahmadvand@shahroodut.ac.ir}
 \author{\large Kazem Bitaghsir Fadafan}
\email{e-mail:bitaghsir@shahroodut.ac.ir}
\affiliation{Physics Department, University of Shahrood, Shahrood,
Iran}


\date{\today}

\begin{abstract}
Using the AdS/CFT correspondence, we probe the extremal black holes by studying the energy loss of a moving
heavy point particle in a strongly-coupled boundary
field theory at zero temperature and finite charge density. We first consider the extremal
Reissner-Nordstr$\rm\ddot{o}$m-AdS (AdSRN) background black hole in the bulk and find an
analytic solution of drag force which depends on the finite charge density of the boundary field theory. By studying the near horizon geometry of trailing string in the bulk, we show the IR divergency does not show the logarithmic behavior. We study also the stochastic behavior of the probe and find the Langevin diffusion coefficients. This study is extended to the extremal backgrounds with vanishing entropy density in appropriate extremal limits.

\end{abstract}
\maketitle
\section{Introduction}
Strongly coupled systems describe interesting phenomena in different areas of physics. Study of such systems is theoretically difficult, then one should develop alternative theoretical tools to understand them better. In the framework of gauge/gravity correspondence, such systems with a large number of degrees of freedom could be realized holographically as classical gravity in a higher dimension. Besides, from this correspondence to study the field theory states at nonzero temperature and density, one should consider geometries with a black brane horizon. The equilibrium thermodynamic properties of such states are described by black hole thermodynamics.

In this paper, we consider zero temperature and finite density systems which correspond to extremal black branes. They have vanishing surface gravity and Hawking temperature. Also given a set of charges, they have the lowest possible mass. Therefore, they can be thought of as ground states of the non extremal black holes and one finds that all supersymmetric BPS black holes are necessarily extremal in supersymmetric gravity theories. Because of the relation between horizon area and entropy, such extremal black branes are dual to field theory states with zero temperature and a large nonzero entropy density. This behavior is non-trivial and somewhat problematic to understand because from third low of thermodynamics, one expects zero entropy at zero temperature. Based on these reasons, extremal black holes have been studied extensively, for example in the case of quantum properties of black holes and microstate counting. Another example of such applications is studying the quantum critical phenomena in condensed matter systems \cite{Faulkner:2009wj}. They studied the geometry of extremal Anti-de Sitter Reissner-Nordstrom (AdSRN) in $d+1$ black brane where the dual field theory is supposed to be a $d$-dimensional strongly-coupled system at zero temperature and finite $U(1)$ charge density. To probe this system, they studied scalars as well as spinors. Some transport coefficients such as conductivity and viscosity have been studied using $AdS_4RN$. It was shown that the ratio of shear viscosity $(\eta)$ to the entropy density $(s)$ is $\frac{1}{4\pi}$ \cite{Edalati:2009bi,Chakrabarti:2009ht,Cai:2009zn,Paulos:2009yk}. As a result, one may generalize some universal arguments at finite temperature to field theory at zero temperature using extremal backgrounds.

The extremal backgrounds with vanishing entropy density in appropriate extremal limits are interesting research area. The extremal black holes with a vanishing one-cycle on the horizon and with vanishing horizon area have been studied in \cite{Sadeghian:2015laa}. The different aspects of a given Extremal Vanishing Horizon (EVH) black hole, its near-horizon geometry and a possible dual two dimension conformal field theory (CFT) picture for the corresponding excitations have been studied in \cite{SheikhJabbaria:2011gc,Johnstone:2013ioa}. They also discuss the EVH/CFT propsal in various theories and diverse dimensions. Moreover, the laws of near horizon extremal geometries dynamics were discussed in \cite{Hajian:2013lna,Hajian:2014twa}. Using "top-down" construction, the four and five dimensional extremal black hole have been studied in \cite{DeWolfe:2013uba,DeWolfe:2014ifa,Cosnier-Horeau:2014qya}. They consider extremal black branes with a singular horizon and zero ground state entropy which are dual to zero-temperature $\mathcal{N} = 4$ Super-Yang-Mills theory with two equal nonvanishing chemical potentials. The singularities are related to divergences of the scalars as well as of curvature invariants and they can be avoided by turning on small but non-zero horizon radius $(r_h)$. One should notice that in the infrared region (IR) of such geometries there is no $AdS_2$ region which is interesting because the $AdS_2$ regions in the extremal limit of more general black holes like AdSRN are responsible for many universal properties of the boundary field theory at zero temperature.

To study field theory properties of such systems, one may use the linear response of the theory to fermionic operators as \cite{DeWolfe:2013uba}. Then Fermi surface singularities and the spectrum of fluctuations should be found and finally classified different behavior as Fermi liquid, non-Fermi liquid, or marginal Fermi liquid. Results can be related to the experimental observations like strange metals in cuprates and in heavy fermion systems \cite{Hartnoll:2009ns}.

In this paper we probe the extremal black holes by studying energy loss of a moving heavy point particle. In the case of studying quark-gluon-plasma (QGP) produced at RHIC and LHC, the energy loss of heavy probes is an important subject \cite{Matsui:1986dk}. Studying such phenomena at strong coupling needs non-perturbative and time dependent methods which are inadequate for describing \cite{CasalderreySolana:2011us,DeWolfe:2013cua}. Using gauge/gravity duality, the energy loss of heavy quarks was initially investigated in \cite{Herzog:2006gh,Gubser:2006bz}. They considered a heavy probe quark traveling with constant linear velocity through
the strongly coupled plasma, and found the energy required to keep it in uniform motion. One finds that the energy loss is proportional to the momentum of the quark and concludes that the energy loss mechanism is such as a drag force.

Study of the energy loss of heavy probes at zero temperature has been done in the case of a strongly coupled superfluids in \cite{Gubser:2009qf}. They use gauge/gravity correspondence and model a superfluid to study its response to a heavy pointlike probe. It was shown that there
is a critical velocity which was determined from the bulk geometry. They also study the stochastic force acting on the moving heavy probe and found the transverse $(\kappa_T)$ and longitudinal $(\kappa_L)$ mean momentum transfer. The extension of the ideas of using AdS/CFT to study energy loss of extended defects in strongly coupled systems has been done in \cite{FuiniJohnF:2011aa}. They consider bulk gravities created by general $Dp$-branes as well as holographic superfluids. For the superfluid case, it was confirmed that there is a cutoff velocity, below which the probe experiences no drag force.

In this paper we consider extremal black holes to study the drag force experiences by the heavy probe. We also study Brownian motion and find the $\kappa_T$ and $\kappa_L$ mean momentum transfer at zero temperature and finite density. We check if they satisfy universal behavior $\kappa_L>\kappa_T$. We show that the world-sheet temperature of trailing string $(T_{ws})$ in the extremal backgrounds is finite while the boundary field theory temperature $(T)$ is zero. We compare $T_{ws}$ with the world-sheet temperature of trailing string in the non-extremal backgrounds and find that it could be larger. In addition, the shape of trailing string in near horizon geometry is studied and it is shown that the IR divergency of extremal black hole would not be logarithmic case.

Curiously, we extend our study to the extremal backgrounds with vanishing entropy density in appropriate extremal limits. We first consider 5-dimensional extremal 3-charge black holes denoted as (2+1)QBH when two of charges are equal. In this case there are two different extremal limits for vanishing temperature and entropy. Interestingly, we find an analytic result for the drag force and it is shown that it does not change in these limits. We also study $\kappa_T$ and $\kappa_L$.

This paper is organized as follows. In the next section, we will consider AdSRN extremal geometry in five dimensions and
study the drag force at zero temperature. We also study in this section the quasi normal modes and the stochastic behavior of the point particle. We extend our results to the case of $\mathcal{N}=4$ SYM theory with three different charge black hole (3QBH) in section three. In section four, we consider 10 dimensional 3QBH and study the same physics. In the last section we summarize our results. %
\section{Energy loss from extremal AdSRN}
In this section we study the drag force felt by a moving heavy probe in the gauge theory at zero temperature and finite density. From the AdS/CFT correspondence, the probe is the end of the string which hangs from boundary to the AdSRN background. The RN black hole with negative cosmological constant is given by the following action \cite{Faulkner:2009wj}
\begin{equation}\label{1}
S=-\frac{1}{2\kappa^2}\int d^5x \sqrt{-g}\bigg(R+\frac{12}{L^2}-\frac{L^2}{g_F^2}F_{\mu\nu}F^{\mu\nu}\bigg),
\end{equation}
here $ \kappa^2=8\pi G $ , $ L $ is AdS radius, $ g_F $ is an effective dimensionless gauge coupling constant, and $ F_{\mu\nu}=\partial_{\mu}A_{\nu}-\partial_{\nu}A_{\mu} $ where $ A_{\mu} $ is $ U(1) $ gauge field. The 5-dimensional solution of equation of motion coming from this action can be obtained as $ AdS_5 $ RN black hole
\begin{equation}\label{RN}
ds^2=\frac{r^2}{L^2}(-f dt^2+d\vec{x}^2)+\frac{L^2}{r^2 f}dr^2,
\end{equation}
here
\begin{equation}\label{3}
f=1-\frac{m}{r^4}+\frac{q^2}{r^6},\,\,\,\,\,\, A_t=\mu
\left(1-\frac{r_h^{2}}{r^{2}}\right)
\end{equation}
and $ t $, $ \vec{x} $ are time and space coordinates within the boundary and $ r $ denotes the radial direction. We can rewrite $ f(r) $ as follows, using $ f(r_h)=0 $ where $ r_h $ denotes the event horizon,
\begin{equation}\label{4}
 f=1-\frac{r_h^4}{r^4}+\frac{q^2}{r^6r_h^2}(r_h^2-r^2).
\end{equation}
The parameters $ m $ and $ q $ are the black hole mass and charge, respectively, and are given as
\begin{equation}\label{5}
m=r_h^4+\frac{q^2}{r_h^2}, \hspace{1cm} q^2=\mu ^2n^2r_h^4,
\end{equation}
here $ n^2=4L^4/3g_F^2 $. The Hawking temperature and entropy of black hole are gained as
\begin{equation}\label{6}
T=\frac{r_h}{\pi L^2}(1-\frac{q^2}{2r_h^6}),~~~~~~~~ S=\frac{1}{4G}(\frac{r_h}{L})^3.
\end{equation}
In the case of $ q^2=2r_h^6 $, the temperature vanishes. As it is clear in the extremal case, the Hawking temperature vanishes while the horizon area is finite. This is one of the peculiar properties of the RN black holes. It would be interpreted as a finite ground state degeneracy which is unusual in the dual field theory.  We focus on these systems and study the drag force.
\subsection{Finding drag force from holography}
In the boundary field theory side, an external heavy probe can be described by a classical string that has
a single end point at the boundary and extends down to the horizon. Because of the non-zero finite density, one expects a non-zero drag force. The trailing string
dynamics is captured by the Nambu-Goto action as follows
\begin{equation}\label{7}
S=-\frac{1}{2\pi \alpha'}\int d\sigma d\tau \sqrt{-\det\, g_{ab}}
\end{equation}
here $ g_{ab} $ is the induced metric on the string world-sheet. The $ (\sigma, \tau) $ coordinates parameterize $ g_{ab} $ where $ a, b $ run over these two dimensions. By choosing $ \tau=t, \sigma=r $ and defining $ \dot{X}=\partial_{\tau}X, X'=\partial_{\sigma}X $ and $\dot{X}.X'=\dot{X}^{\mu}X'^{\nu}G_{\mu\nu}$, where $G_{\mu\nu}$ is the background metric (\ref{RN}), one finds
\begin{equation}\label{8}
-\det\, g=(\dot{X}\cdot X')^2-X'^2\dot{X}^2=1-\frac{\nu ^2}{f}+\frac{r^4}{L^4}x'^2f.
\end{equation}
The string in this case trails behind its boundary endpoint as it moves at constant velocity in the $x$ direction
\be
x(r,t)=\nu t +x(r),\,\,\,x^2=x^3=0.
\ee
The equation of motion leads to
\begin{equation}\label{9}
x'^2=\frac{L^4C^2(1-\nu ^2 f^{-1})}{r^4f(r^4fL^{-4}-C^2)},
\end{equation}
here $ C $ is a constant of motion. To have a hanging string from boundary to horizon, $ x'^2 $ should be positive everywhere, i.e. both numerator and denominator of (\ref{9}) should change sign at the same point. This critical point is denoted as $ r_c $ and one finds the constant of motion as
\begin{equation}\label{10}
C=\nu r_c^2 L^{-2}.
\end{equation}
The drag force experienced by the heavy quark is calculated by the following formula
\begin{equation}\label{11}
F=-\frac{1}{2\pi \alpha '}\frac{\partial\mathcal{L}}{\partial x'}=-\frac{1}{2\pi \alpha '}\frac{\nu r_c^2}{L^2}.
\end{equation}
By changing variables, $ \beta ^{-2}=1-\nu^2, y=r_c^2 $, we try to find the $r_c$, by solving $ f(r_c)-\nu ^2=0 $. One finds this equation in terms of the new variables in the non-extremal and extremal cases as follows:
\itemize{
\item{
Non-extremal:
\begin{equation}\label{12}
 y^3-\beta ^2 (r_h^4+\frac{q^2}{r_h^2})y+\beta ^2 q^2=0.
\end{equation}}

\item{Extremal:
\begin{equation}\label{13}
y^3-3r_h^4\beta ^2 y+2r_h^6\beta ^2=0.
\end{equation}}}
The solution of (\ref{12}) can be found in \cite{Fadafan:2008uv}, which the effects of charge and finite 't Hooft coupling correction on the drag force are investigated. In the extremal equation (\ref{12}), we gain three real roots which can be written in the following form such that the range of velocity is $ 0<\beta ^{-1}<1 $
\begin{eqnarray}\label{14}
y_1&=&2r_h^2\beta \cos(\frac{\theta}{3}),\\
y_2&=&2r_h^2\beta \cos(\frac{2\pi +\theta}{3}),\\
y_3&=&2r_h^2 \beta \cos(\frac{4\pi +\theta}{3}),
\end{eqnarray}
here $ \theta =\cos ^{-1}(-\beta ^{-1}) $.

By studying these roots one concludes that $ y_2 $ is not physically valid because $ -\beta ^{-1}<0 $, hence,  $ \pi/6<\theta/3 <\pi/2 $, which $ y_2=r_c^2 $ is proportioned to $ \cos (\theta /3 +2\pi /3) $. Also, $ y_1 $ and $ y_3 $ give identical outcomes. Notice that for the non-extremal case, by setting $ q $ and $ r_h $ parameters fixed one can also come up with the solutions like those of the extremal case.

\begin{figure}[th]
\includegraphics[scale=.8]{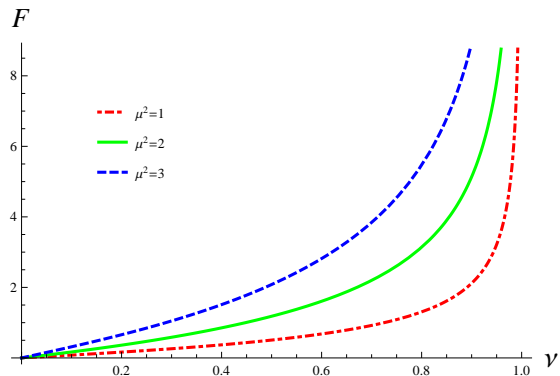}\caption{\label{fo} Drag force is ploted as $ r_h=1 $, $ L=1 $, $ n=1 $ and $ \lambda =100 $ for extremal, $ \mu ^2=2 $ (solid line), and non-extremal case, $ \mu ^2=1 $ (dotdashed line), and $ \mu ^2=3 $ (dashed line).}
\end{figure}

Eventually, the drag force for the extremal case of this background would be
\be \label{17} F=-\frac{\sqrt{\lambda}}{3 g_F^2\pi }\frac{2 \nu \mu ^2}{\sqrt{1-\nu^2}}\cos(\frac{\theta}{3})\ee%
which $ 2r_h^2=\mu^2n^2 $ and $ \alpha'=L^2/\sqrt{\lambda} $ are used where $ \lambda $ is 't Hooft coupling. One can expand this equation for small velocities and it turns out the drag force is proportional to $ \nu $ and $ \nu^3 $ for first and second order of the expansion, respectively. One finds from Fig. \ref{fo} the drag force versus the velocity. We assume $ r_h=1 $, $ L=1 $, $ n=1 $ and $ \lambda =100 $ and consider different values for the chemical potential $\mu$ as $\mu ^2=1,2,3$. As one expects from (\ref{17}), at constant velocity increasing the chemical potential leads to increasing of the drag force.

\subsection{Quasinormal modes}
Now, we consider decelerating of the probe at zero temperature and finite density. We study curved open string on the late time and
low velocity, it means that we apply small perturbations to the classical open string and consider quasinormal mode on the
string worldsheet. These modes give important information about the return to equilibrium of the trailing string after small
perturbations. Therefore, one finds the friction coefficient term, $\tilde{\mu}$, in the small velocity limit. The quasinormal
mode analysis in three dimensions and finite temperature is studied in \cite{Herzog:2006gh}.

In the linear analysis, $\dot{x}$ and $x'$
are very small and one can consider only linear terms in the
equation of motion of string. Assuming time dependent solution as $x(t)=e^{-\tilde{\mu} t}$, the equation of motion becomes%
\begin{equation}
f(r)\partial_r\left( \frac{r^4}{L^4}\,f(r)\,x' \right)
=\tilde{\mu}^2 x.
\end{equation}
In the case of extremal black holes, one can expand $ f(r) $  close to the horizon as $ f''(r_h) (r-r_h)^2+\mathcal{O}(r-r_h)^3 $ where we can set $ n^2 $ such that $ f''(r_h)=\mu ^{-2} $ and obtain the solution as
\begin{equation}
x=c_1~ e^{\frac{\mu ^2 \tilde{\mu}}{r_h^2(r-r_h)}}+c_2~ e^{-\frac{\mu ^2 \tilde{\mu}}{r_h^2(r-r_h)}}\label{xsol}
\end{equation}
which due to out-going boundary conditions we take the first part as the solution \cite{Son:2002sd}. It would be interesting to compare (\ref{xsol}) with non-extremal case in \cite{Herzog:2006se} as
\be
x_N=a_1\left(r-r_h\right)^{\frac{\tilde{\mu}}{4\pi T}}+a_2\left(r-r_h\right)^{-\frac{\tilde{\mu}}{4\pi T}}.
\ee
As seen, two solutions are not the same. By expanding $ x $ in (\ref{xsol}), we arrive at
\begin{equation}
x=c_1~(1+\frac{\mu ^2 \tilde{\mu}}{r_h^2(r-r_h)}+\mathcal{O}(\mu ^4)).
\end{equation}
Therefore, taking derivative, it yields $ x' \approx -\tilde{\mu} \mu ^2/(r_h^2(r-r_h)^2) $ while for the non-extremal case the shape of the string is obtained as $ x_N' \approx -\tilde{\mu}/(4\pi T(r-r_h)) $. As shown, the near horizon behavior of trailing string in extremal black holes differs from that of the non-extremal case. In the former it  is proportional to $(r-r_h)^{-1}$ while in the latter it has logarithmic divergency. By studying the near horizon geometry of trailing string in the bulk, one finds that the IR divergency does not show the logarithmic behavior in the extremal case.
We show in the ten dimensional case that this divergency depends on the charges of the background.

In general one may also consider an exponential damping solution of the form,
\begin{equation}
x(r,t)=\mathcal{A}(r) e^{-\tilde{\mu} t}.
\end{equation}
which means that $\dot{x}=-\tilde{\mu}\,x$. The equation of motion becomes the eigenvalue equation
\begin{equation}
\mathcal{O}\,\mathcal{A}(r)=\tilde{\mu}^2\,\mathcal{A}(r),\,\,\,\,\,\,\,\
\mathcal{O}= f(r)\frac{d}{dr}\left(r^4\,f(r)\,\frac{d}{dr}\right)\label{oper}.
\end{equation}
We have chosen $L=1$. For small $\tilde{\mu}$, we expand $x$ as a power series $x=x_0+\tilde{\mu}^2 x_1+\cdots $. The eigenvalue equation implies that
\be \mathcal{O}x=\tilde{\mu}^2x.\label{eigen}\ee

As a result, $\mathcal{O}x_0=0$ and $\mathcal{O}x_1=x_0$.
The general linear analysis and some universal features of studying quasi normal modes have been studied in \cite{Herzog:2006se}. However, for extremal black holes the near horizon behavior is not the same as the non-extremal case. If we consider the moving particle on a flavor brane located at the bulk radius $r=r_0$, for large $ r_0 $ and near $ r=r_0 $, $ f $ tends to 1. Therefore, by taking $ x_0=A $, which is a constant, and using (\ref{oper}) and (\ref{eigen}), one finds for $ r\sim r_h $
\begin{equation}
x'\approx \frac{-A \mu ^2\tilde{\mu} ^2 }{r_h^4 (r-r_h)^2}\bigg( r_0+\frac{\mu ^2}{(r-r_h)} \bigg).
\end{equation}
Hence we can find the friction coefficient as
\begin{equation}
\tilde{\mu}=\frac{r_h^2}{r_0}
\end{equation}
so that the momentum flowing down the string is
\begin{equation}
\frac{dp}{dt}=-\tilde{\mu}p=-\frac{1}{2\pi \alpha '}\nu r_h^2
\end{equation}
where $ m=r_0/2\pi \alpha ' $ for large $ r_0 $ limit. This is exactly the result of (\ref{11}) if we expand $ r_c $ around $ r_h $ for small velocities.
\subsection{Stochastic behavior}
\begin{figure}[th]
\includegraphics[scale=.8]{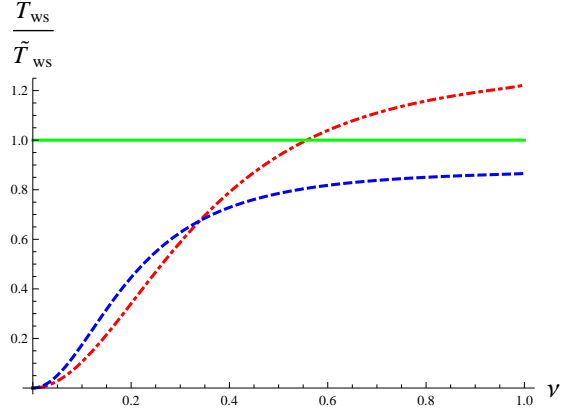}\caption{\label{f} $ T_{ws} $ is world-sheet temperature of the extremal case and $ \tilde{T}_{ws} $ is that of the non-extremal case. We plot this ratio for $ r_h=1 $ and $ \mu ^2=1$ (dashed), $ \mu ^2=2 $ (solid line) which is extremal, and $ \mu ^2=3 $ (dotdashed).}
\end{figure}

We also calculate stochastic forces on the moving quark with constant velocity. These forces depend on the existence of the horizon on the string world-sheet which is identical with critical point, $ r_c $. This motion which is analogous to the Brownian motion may be found by the generalized Langevin equations involving Langevin coefficients which are proportional to the temperature of the string world-sheet. This temperature is given by the following formula \cite{Giataganas:2013zaa}
\begin{eqnarray}\label{19}
T_{ws}&=&\frac{1}{16\pi^2}\Bigg | {\frac{1}{g_{tt}g_{rr}}(g_{tt}g_{xx})'(\frac{g_{tt}}{g_{xx}})' }\Bigg | \bigg |_{r=r_c}\nonumber \\&=&\frac{3 \mu ^4 n^4 \sec ^5\left(\frac{\theta }{3}\right) \bigg(\mu ^2 n^2-4 \beta  \cos \left(\frac{\theta }{3}\right)\bigg)^2 \bigg(16 \beta ^2 \cos ^2\left(\frac{\theta }{3}\right)+\mu ^4 n^4+4 \beta  \mu ^2 n^2 \cos \left(\frac{\theta }{3}\right)\bigg)}{2048 \pi ^2 \beta ^5 L^4}
\end{eqnarray}
We computed this temperature in the extremal case and as seen from (\ref{6}) and (\ref{14}) in finite temperature case the place of $ r_c $ and as a result $ T_{ws} $ would change. As seen from Fig. \ref{f}, we compared this quantity in extremal and non-extremal cases for different values of $ \mu $ with $ r_h=1 $. As velocity rises, the ratio of world-sheet temperature in the extremal case to non-extremal world-sheet temperature increases. Furthermore, we must note world-sheet temperature of the extremal case becomes more than that of the non-extremal with $ \mu ^2=1 $ for $ \nu >0.55 $ whereas this ratio is less than one for $ \mu ^2=3 $ and all velocities. \\\\

\begin{figure}[th]
\includegraphics[scale=.8]{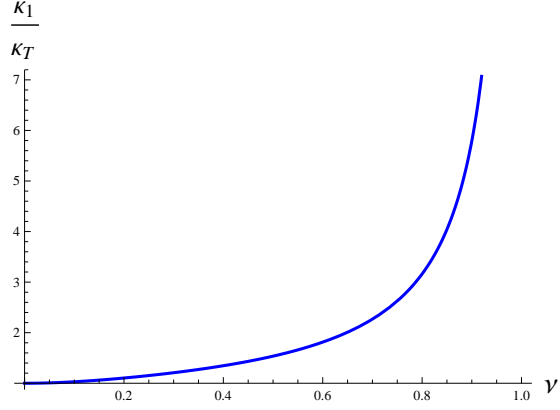}\caption{\label{ff} $ \kappa_{L}/\kappa_{T} $ ratio is shown for the extremal case.}
\end{figure}
The Langevin coefficients also can be achieved from the background metric components
\begin{equation}\label{20}
\kappa_{T}=\frac{1}{\pi \alpha'}g_{kk}\bigg |_{r_c} T_{ws}=\frac{3 \mu ^4 n^4\sqrt{\lambda} \bigg(\mu ^2 n^2 \sec \left(\frac{\theta }{3}\right)-4\beta \bigg)^2 \bigg(16 \beta ^2+\mu ^4 n^4 \sec ^2\left(\frac{\theta}{3}\right)+4 \beta  \mu ^2 n^2 \sec \left(\frac{\theta}{3}\right)\bigg)}{1024 \pi ^3 \beta ^4 L^8}
\end{equation}
where the index $ k $ in $g_{kk}$ denotes a particular transverse direction of motion $ x $, which is $ x^2 $ or $ x^3 $ coordinates, and for the longitudinal direction it is
\begin{eqnarray}\label{21}
\kappa_{L}&=&\frac{1}{\pi\alpha'}\frac{(g_{tt}g_{xx})'}{g_{xx}(\frac{g_{tt}}{g_{xx}})'}\bigg |_{r=r_c} T_{ws}\nonumber \\&=&\frac{\sqrt{\lambda}}{1024 \pi ^3 \beta ^4 L^8}\bigg(16 \beta ^2 \cos ^2(\frac{\theta }{3})+\mu ^4 n^4+4\beta  \mu ^2 n^2 \cos (\frac{\theta }{3})\bigg) \bigg(\mu ^2 n^2 \sec (\frac{\theta }{3})-4 \beta \bigg)^2\nonumber \\ &\times &\bigg(16\beta ^2+\mu ^4 n^4 \sec ^2(\frac{\theta }{3})+4 \beta  \mu ^2 n^2 \sec (\frac{\theta }{3})\bigg)
\end{eqnarray}
In Fig. \ref{ff}, we plot $ \kappa_{L}/\kappa_{T} $ quantity in terms of velocity for the extremal case, $ \mu ^2=2 $, as $ r_h=1 $. It is indicated these coefficients satisfy the inequality $ \kappa_{L}>\kappa_{T} $, and it also holds for the non-extremal case, \cite{Giataganas:2013hwa}.

\section{Energy loss from 5-dimensional extremal 3-charge black holes}
\begin{figure}[th]
\includegraphics[scale=.8]{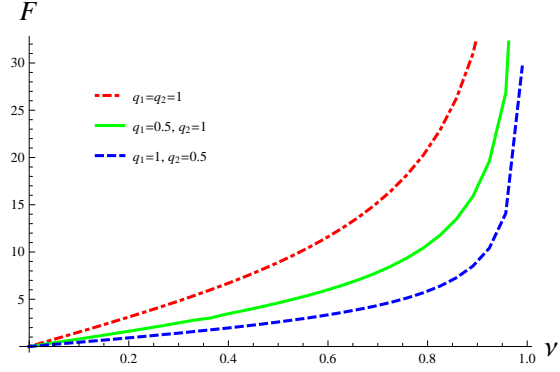}\caption{\label{f0} We plot non-extremal (2+1)QBH with parameter space for $ r_h=0.1 $ as $ q_1= 1$ $ (\mu _1=2.5)$, $ q_2=0.5 $ $ (\mu _2=7.1)$, dashed, $ q_1=0.5 ~ (\mu _1=9.9), q_2=1 ~ (\mu _2=7.2)$, line, and $ q_1=1 ~ (\mu _1=10), q_2=1 ~ (\mu _2=14.2) $, dotdashed.}
\end{figure}

$ \mathcal{N}=4 $ SYM theory is dual to type IIB supergravity on $ AdS_5\times S^5 $ which has a rank 3 SO(6) R-symmetry associated to $ S^5 $ and corresponded to 3 distinct chemical potentials. Indeed, $ \mathcal{N}=4 $ SYM theory at nonzero temperature and chemical potential is dual to the near horizon limit of rotating D3-background consisting of a Kerr black hole with $ AdS_5\times S^5 $ asymptotic. The theory has three different charges refereed to as 3-charge black hole (3QBH). For simplicity we can take two of them equal, i.e. (2+1)QBH which in this case if $ q_1=0 $, it is called 2QBH.
The effective supergravity Lagrangian is given by \cite{DeWolfe:2013uba}
\begin{equation}\label{22}
e^{-1}\mathcal{L}=R-\frac{1}{2}(\partial\phi)^2+\frac{8}{L^2}e^{\frac{\phi}{\sqrt{6}}}+\frac{4}{L^2}e^{\frac{-2\phi}{\sqrt{6}}}-e^{\frac{-4\phi}{\sqrt{6}}}f_{\mu\nu}f^{\mu\nu}-2e^{\frac{2\phi}{\sqrt{6}}}F_{\mu\nu}F^{\mu\nu}-2\epsilon^{\mu\nu\rho\sigma\tau}f_{\mu\nu}F_{\rho\sigma}A_{\tau}
\end{equation}
where $ A_{\mu} $ is the gauge field related to the two equal charges and $ a_{\mu} $ is associated with the other charge. Five-dimensional black brane solution of (\ref{22}) yields the metric of background
\begin{equation}\label{23}
ds^2= e^{2A(r)}(-h(r)dt^2+d\vec{x}^2)+\frac{e^{2B(r)}}{h(r)}dr^2
\end{equation}
where
\begin{eqnarray}\label{24}
A(r)&=&\log \frac{r}{L}+\frac{1}{6}\log (1+\frac{q_1^2}{r^2})+\frac{1}{3}\log (1+\frac{q_2^2}{r^2}),\\
B(r)&=&-\log \frac{r}{L}-\frac{1}{3}\log (1+\frac{q_1^2}{r^2})-\frac{2}{3}\log (1+\frac{q_2^2}{r^2}) ,\\
h(r)&=&1-\frac{r^2(r_h^2+q_1^2)(r_h^2+q_2^2)^2}{r_h^2(r^2+q_1^2)(r^2+q_2^2)^2}.
\end{eqnarray}
Temperature, entropy and chemical potentials are given by the following relations
\begin{eqnarray}\label{25}
T&=&\frac{2r_h^4+q_1^2r_h^2-q_1^2q_2^2}{2\pi L^2 r_h^2\sqrt{r_h^2+q_1^2}},~~~~~S=\frac{1}{4GL^3}\sqrt{r_h^2+q_1^2}(r_h^2+q_2^2),\\
\mu _1&=&\frac{q_1(r_h^2+q_2^2)}{L^2 r_h\sqrt{r_h^2+q_1^2}},~~~~\mu _2=\frac{\sqrt{2}q_2\sqrt{r_h^2+q_1^2}}{L^2 r_h}.
\end{eqnarray}
Defining $ X=(t, r, x) $, since we assume the stationary motion of the heavy quark, from equation of motion we can gain
\begin{equation}\label{26}
x'^2=\frac{[e^{2(A+B)}(1-\frac{\nu^2}{h(r)})]\Pi ^2}{(h(r)e^{4A}-\Pi ^2)h(r)e^{4A}}
\end{equation}
where $ \Pi $ is a constant of motion. Due to changing sign of numerator and denominator at the same for positivity of (\ref{26}) everywhere, this constant would be $ \Pi=\nu \exp (2A(r_c)) $. Therefore, the drag force is
\begin{equation}\label{27}
F=-\frac{1}{2\pi \alpha'}\frac{\partial \mathcal{L}}{\partial x'}=-\frac{1}{2\pi \alpha'}\nu e^{2A(r_c)}.
\end{equation}
For non-extremal $(2+1)$ QBH case, using $ \nu^2=h(r_c) $, one can obtain the related drag force. As seen from Fig. \ref{f0}, we ploted the drag force as a function of velocity with various parameters.
\subsection{2QBH extremal limits}
There are two different limits for obtaining extremal 2QBH. For the first approach (type I), one can set $ q_1\rightarrow 0 $, then adjust $ r_h\rightarrow 0 $ for extremality. The second one (type II) is that achieving $ q_1 $ as a function of $ r_h $ and $ q_2 $, then imposing $ r_h \rightarrow 0 $. These two limits are not exactly the same and quantities like chemical potential take different values. We may investigate drag force for extremal 2QBH case obtained through these two limits. To reach extremal 2QBH (type I), we set $ q_1 $ equal to zero which turns non-extremal 2QBH. Then, For this extremal 2QBH, $ q_1\rightarrow 0 $
\begin{equation}\label{28}
A=\log\frac{r}{L}+\frac{1}{3}\log(1+\frac{q_2^2}{r^2}),~~~~~~~~~ B=-\log\frac{r}{L}-\frac{2}{3}\log(1+\frac{q_2^2}{r^2}),
\end{equation}
\begin{equation}\label{29}
h=1-\frac{q_2^4}{(r^2+q_2^2)^2},~~~~~~~~S\rightarrow 0 ,~~~~~~~~\mu _1\rightarrow 0,~~~~~~~~\mu _2=\frac{\sqrt{2}q_2}{L^2},~~~~~~~~r_c^2=q_2^2(-1+\frac{1}{\sqrt{1-\nu ^2}}).
\end{equation}
where $ r_c $ is found from $ h(r_c)=\nu^2 $ for $ 0<\nu<1 $. Eventually, drag force for this case would be
\begin{equation}\label{30}
F=-\frac{\nu}{2\pi \alpha' L^2}\frac{q_2^2(1-\sqrt{1-\nu ^2})^{1/3}}{\sqrt{1-\nu ^2}}= -\frac{\nu \mu _2^2 \sqrt{\lambda}}{4\pi }\frac{ (1-\sqrt{1-\nu ^2})^{1/3}}{\sqrt{1-\nu ^2}}.
\end{equation}
For extremal 2QBH (type II), using $ T=0 $, we obtain $ q_1^2=2r_h^4/(q_2^2-r_h^2) $ which is extremal (2+1) QBH case. By plugging $ q_1^2 $ into the relevant equations we find
\begin{eqnarray}\label{31}
A=\log\frac{r}{L}+\frac{1}{6}\log(1+\frac{2r_h^4}{q_2^2-r^2}) +\frac{1}{3}\log(1+\frac{q_2^2}{r^2}),~~~~~~~B=-\log\frac{r}{L}-\frac{1}{3}\log(1+\frac{2r_h^4}{q_2^2-r^2}) -\frac{2}{3}\log(1+\frac{q_2^2}{r^2})
\end{eqnarray}
\begin{eqnarray}
h=1-\frac{r^2(1+\frac{2r_h^2}{q_2^2-r_h^2})(r_h^2+q_2^2)^2}{(r^2+\frac{2r_h^4}{q_2^2-r_h^2})(r^2+q_2^2)^2},~~~~~~~S=\frac{1}{4GL^3}\frac{r_h^2(q_2^2+r_h^2)^{3/2}}{(q^2_2-r_h^2)^{1/2}},
\end{eqnarray}
\begin{equation}
 \mu_1=\frac{\sqrt{2}(r_h^2+q_2^2)}{L^2\sqrt{(q_2^2-r_h^2)(1+\frac{2r_h^2}{q^2_2-r_h^2})}},~~~~~~~\mu_2=\frac{\sqrt{2}q_2\sqrt{1+\frac{2r_h^2}{q_2^2-r_h^2}}}{L^2}.
\end{equation}
Then, by setting $ r_h \rightarrow 0 $, we achieve the requested case. In (type II) case, one may gain $ A, B $, and $ r_c $ the same as the ones (type I) except for $ \mu _1=\mu _2=\sqrt{2}q_2/L^2 $. Thus, drag force for extremal 2QBH (type II) would be
\begin{equation}\label{33}
F=-\frac{\nu}{2\pi \alpha' L^2}\frac{q_2^2 (1-\sqrt{1-\nu ^2})^{1/3}}{\sqrt{1-\nu ^2}}= -\frac{\nu \mu _1^2 \sqrt{\lambda}}{4\pi }\frac{(1-\sqrt{1-\nu ^2})^{1/3}}{\sqrt{1-\nu ^2}}
\end{equation}
In these two limits, $ h(r_c) $ can be expressed in terms of $ q_2 $ associated with $ \mu _2 $ which is equivalent for both. Therefore, the drag force for Extremal 2QBH cannot be affected from two limits. However, in type II limit, there is the other nonzero chemical potential, $ \mu_1 $, in terms of which the drag force may be expressed. ( See \cite{Sadeghi:2009hh} for drag force in non-extremal cases with different charges.) Also, as seen from (\ref{26}), the shape of string in the bulk for these two limits would be the same since A and B as well as $ r_c $ give equal results for the limits.\\

\subsection{Stochastic behavior}
From (\ref{23}), we can obtain the temperature of the word-sheet in this background
\begin{eqnarray}\label{37}
T_{ws}&=&\frac{1}{16\pi^2}\Bigg (e^{2(A(r)-B(r))}h'(r)(4A'(r)+h'(r)) \Bigg)\bigg |_{r=r_c}\nonumber \\&=&\frac{q_2^4}{3 \pi ^2 L^4 (\beta +q^2_2-1)^4}\bigg(3 (\beta -1)^3+7q_2^2 (\beta -1)^2+8q_2^4 (\beta -1)+q_2^6\bigg)
\end{eqnarray}
where $ r_c^2=q_2^2 (\beta -1) $ and is the same for these two extremal limits. Langevin coefficients can be achieved as well. For the transverse direction, this coefficient can be formulated as
\begin{eqnarray}\label{38}
\kappa_{T}&=&\frac{1}{\pi\alpha'}e^{2A(r_c)}T_{ws}\nonumber \\&=&\frac{q_2^4\sqrt{\lambda}}{3 \pi ^3 L^8(\beta +q_2^2-1)^4}\bigg[\bigg(\frac{1}{\beta -1}+1\bigg)^{2/3}\bigg(3 (\beta -1)^3+7q_2^2 (\beta -1)^2+8q_2^4 (\beta -1)+q_2^6\bigg)\bigg(\beta -1\bigg)\bigg]
\end{eqnarray}
and for the longitudinal one
\begin{eqnarray}\label{39}
\kappa_{L}&=&\frac{1}{\pi\alpha'}\frac{e^{2A(r_c)}(4A'(r_c)+h'(r_c))}{h(r_c)} T_{ws}\nonumber \\&=&\frac{4q_2^4\sqrt{\lambda}}{9\pi ^3L^8}\frac{\bigg(\frac{1}{\beta -1}+1\bigg)^{2/3}\bigg(3 (\beta -1)^3+7q_2^2 (\beta -1)^2+8q_2^4 (\beta -1)+q_2^6\bigg)^2}{\sqrt{\beta -1} (\beta +q_2^2-1)^5 (\beta +2q_2^2-1)}.
\end{eqnarray}
\begin{figure}[th]
\includegraphics[scale=.8]{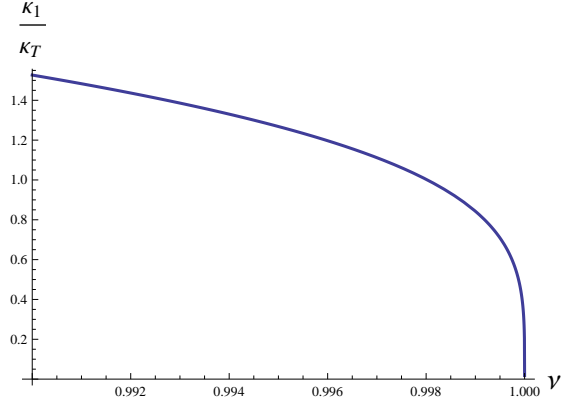}\caption{\label{flan}For the Extremal 2QBH, $ \kappa_{L}/\kappa_{T} $ vesus high velocity is shown as $ \mu _2=\mu _1=\sqrt{2} $, and $ r_h=0 $.}
\end{figure}
Notice the ratio of langitudinal to transverse coefficient is not more than one for the whole range of velocities and it is less than one for ultra-relativistic velocities, Fig. \ref{flan}.
\section{Energy loss from 10-dimensional lift extremal 3QBH}
In this section, we consider ten-dimensional lift of 3-charge black holes as solutions of type IIB supergravity and study different extremal cases in this background. Study of trailing strings in this geometry and at finite temperature and R-charged chemical potential has been done in \cite{Herzog:2007kh}. They find a large number of solutions in this background and show that the single charge solution of \cite{Caceres:2006dj} belongs to them. The metric is given by \cite{Cvetic:1999xp} as
\begin{eqnarray}\label{47}
ds^2&=& \sqrt{\Delta}\left(-(H_1H_2H_3)^{-2/3}h dt^2+(H_1H_2H_3)^{1/3}[h^{-1}dr^2+r^2(d\chi_1^2+\sin ^2 \chi_1(d\chi_2^2+\sin ^2\chi_2 d\chi_3^2))]\right)\nonumber \\ &+&\frac{L^2}{\sqrt{\Delta}}\sum_{i=1}^3 H_i(H_1H_2H_3)^{-1/3}(d\varphi_i^2+\varphi_i^2(d\psi_i+j^i dt)^2)
\end{eqnarray}
where above functions are
\begin{equation}\label{48}
\Delta=\sum_{i=1}^3 H_i(H_1H_2H_3)^{1/3}\varphi _i^2,~~~~~ h=1- \frac{m}{r^2}+\frac{r^2}{L^2}H_1H_2H_3,~~~~~ H_i=1+\frac{q_i}{r^2},~~~~~ j^i=\frac{(1-H_i^{-1})\sqrt{q_i(m+q_i)}}{L q_i},
\end{equation}
\begin{equation}\label{49}
\varphi_1=\sin\theta_1,~~~~~~ \varphi_2=\cos\theta_1\sin\theta_2,~~~~~~ \varphi_3=\cos\theta_1\cos\theta_2.
\end{equation}
$ \chi _i $ are coordinates on 3-sphere and $ S^5 $ is parameterized by $ (\theta_1, \theta_2, \psi_1, \psi_2, \psi_3) $. Hawking temperature and entropy of the black hole are
\begin{equation}\label{50}
T=\frac{1}{2\pi r_h^2L^2}\frac{2r_h^6+r_h^4(q_1+q_2+q_3+L^2)-q_1q_2q_3}{\prod_{i=1}^3 \sqrt{r_h^2+q_i}},
\end{equation}
\begin{equation}\label{51}
S=\frac{\pi ^2}{2G}\prod_{i=1}^3\sqrt{r_h^2+q_i}
\end{equation}
where $ G $ is the five-dimensional Newton's constant and horizon radius, $ r_h $, is obtained from $ h(r_h)=0 $. For simplicity we ignore $ \theta _i $ directions and rewrite the metric as
\begin{equation}\label{52}
ds^2=f[-\alpha dt^2+\beta dr^2+\gamma (d\chi_1^2+\sin ^2 \chi_1(d\chi_2^2+\sin ^2\chi_2 d\chi_3^2))]+\frac{1}{f}\sum_{i=1}^3\kappa_i(d\psi_i+j^idt)^2
\end{equation}
where
\begin{equation}\label{53}
\sqrt{\Delta}\equiv f,~~~~~~~~ h (H_1H_2H_3)^{-2/3}\equiv\alpha, ~~~~~~~~ \frac{(H_1H_2H_3)^{1/3}}{h}\equiv\beta,
\end{equation}
\begin{equation}
 r^2 (H_1H_2H_3)^{1/3}\equiv\gamma,~~~~~~ H_i(H_1H_2H_3)^{-1/3}L^2\equiv\kappa_i.
\end{equation}
We are going to investigate stationary solutions in which $ \vartheta_i= \dot{\chi_i} $ and $ \omega_i = \dot{\psi_i} $ are constant by taking $ \tau=t, \sigma=r $. Defining $ X=(t, r, \chi_1, \chi_2, \chi_3, \psi_1, \psi_2, \psi_3) $, Lagrangian can be obtained as $ \mathcal{L}^2=(X'\cdot\dot{X})^2-X'^2\dot{X}^2 $ where
\begin{eqnarray}\label{54}
X'\cdot\dot{X}&=&f\gamma[\vartheta_1 \chi'_1+\sin ^2 \chi_1(\vartheta _2\chi'_2+\sin ^2\chi_2 \vartheta _3\chi'_3)]+\frac{1}{f}\sum_{i=1}^3 \kappa_i(\omega _i+j^i)\psi '_i\\
X'^2&=&f\beta +f\gamma [(\chi '_1)^2+\sin ^2 \chi _1((\chi '_2)^2+\sin ^2\chi _2 (\chi '_3)^2)]+\frac{1}{f}\sum_{i=1}^3\kappa _i \psi '_i\\
\dot{X}^2&=&-f\alpha +f\gamma[\vartheta_1 ^2+\sin ^2 \chi_1(\vartheta _2^2+\sin ^2\chi_2 \vartheta _3^2)]+\frac{1}{f}\sum_{i=1}^3\kappa_i(\omega_i+j^i)^2 .
\end{eqnarray}
Relevant momenta for the string spining in $ \psi_1=\psi $ and $ \chi _i $ directions corresponded to associated drag forces are gained as
\begin{eqnarray}\label{57}
\Pi_{\chi_1}&=&\frac{-1}{2\pi\alpha'}\frac{\partial\mathcal{L}}{\partial \chi'_1}= \frac{-1}{2\pi\alpha'\mathcal{L}}[f^2\gamma^2\sin^2\chi_1(\vartheta_1\vartheta_2\chi'_2 + \vartheta_1\vartheta_3\chi'_3\sin^2\chi_2)+\gamma\kappa_1(\omega_1+j^1)\vartheta_1\psi'+f^2\alpha\gamma \chi'_1 \nonumber \\&-&f^2\gamma^2\chi'_1\sin ^2 \chi_1(\vartheta _2^2+\sin ^2\chi_2 \vartheta _3^2)-\sum_{i=1}^3\gamma\kappa_i(\omega_i+j^i)^2\chi'_1],\\
\Pi_{\chi_2}&=&\frac{-1}{2\pi\alpha'}\frac{\partial\mathcal{L}}{\partial \chi'_2}= \frac{-1}{2\pi\alpha'\mathcal{L}}[\sin^2\chi_1 (f^2\gamma^2(\vartheta_1\vartheta_2\chi'_1 +\vartheta_2\vartheta_3\chi'_3\sin^2\chi_1\sin^2\chi_2)+\gamma\kappa_1(\omega_1+j^1)\vartheta_2\psi' \nonumber \\&+&f^2\alpha\gamma \chi'_2-f^2\gamma^2\chi'_2(\vartheta_1 ^2+\sin ^2 \chi_1\sin ^2\chi_2 \vartheta _3^2)-\sum_{i=1}^3\gamma\kappa_i(\omega_i+j^i)^2\chi'_2 )],\\
\Pi_{\chi_3}&=&\frac{-1}{2\pi\alpha'}\frac{\partial\mathcal{L}}{\partial \chi'_3}= \frac{-1}{2\pi\alpha'\mathcal{L}}[\sin^2\chi_1\sin^2\chi_2 (f^2\gamma^2(\vartheta_1\vartheta_3\chi'_1 +\vartheta_2\vartheta_3\chi'_2\sin^2\chi_1)+\gamma\kappa_1(\omega_1+j^1)\vartheta_3\psi' \nonumber \\&+&f^2\alpha\gamma \chi'_3 -f^2\gamma^2\chi'_3(\vartheta_1 ^2+\sin ^2 \chi_1\vartheta _2^2)-\sum_{i=1}^3\gamma\kappa_i(\omega_i+j^i)^2\chi'_3 )],\\
\Pi_{\psi_i}&=&\frac{-1}{2\pi\alpha'}\frac{\partial\mathcal{L}}{\partial \psi'}= \frac{-1}{2\pi\alpha'\mathcal{L}}[\sum_{i=1}^3\gamma\kappa_i(\omega_i+j^i)(\vartheta_1 \chi'_1+\sin ^2 \chi_1(\vartheta _2\chi'_2+\sin ^2\chi_2 \vartheta _3\chi'_3))+\alpha\kappa_1\psi'\nonumber \\&-&\gamma\kappa_1\psi'(\vartheta_1 ^2+\sin ^2 \chi_1(\vartheta _2^2+\sin ^2\chi_2 \vartheta _3^2))-\frac{1}{f^2}\kappa_2\kappa_1(\omega_2+j^2)^2\psi'-\frac{1}{f^2}\kappa_3\kappa_1(\omega_3+j^3)^2\psi'].
\end{eqnarray}
We, however, figure out from above equations that these quantities are no longer constant and satisfy their following equation of motion which for these directions they are
\begin{equation}\label{61}
\partial _r (\frac{\partial \mathcal{L}}{\partial X'})+\partial _t (\frac{\partial \mathcal{L}}{\partial \dot{X}})-\frac{\partial \mathcal{L}}{\partial X} =0.
\end{equation}
To avoid this complexity, we focus on the case where $ \chi_i=const $. As a result, the drag force would be
\begin{equation}\label{62}
F_{\psi}=-\frac{1}{2\pi\alpha'}\frac{\partial\mathcal{L}}{\partial \psi'}= -\frac{1}{2\pi\alpha'\mathcal{L}}(\frac{\kappa_1\psi'}{f^2}I_1)
\end{equation}
where $ I_1=\alpha f^2-\kappa_2(j^2)^2-\kappa_3(j^3)^2 $. Plugging $ \mathcal{L} $ into the formula, we have $ (\psi')^2=\Pi^2 I_2/(\frac{\kappa_1^2}{f^4}I_1^2-\frac{\kappa_1(\Pi)^2}{f^2}I_1) $ where $ I_2=\alpha\beta f^2-\beta[\kappa_1(\omega+j^1)^2+\kappa_2(j^2)^2+\kappa_3(j^3)^2] $ and $ \Pi $ is a constant of motion. One should calculate drag force at the critical point, which this leads to the following relations
\begin{figure}[th]
\includegraphics[scale=0.8]{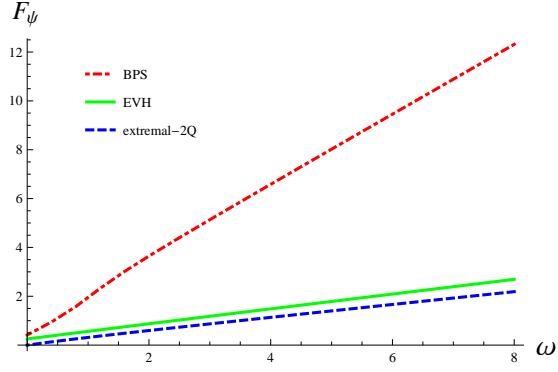}\caption{\label{f3}for $ \lambda =100 $, BPS case with  $ q_1=1, q_2=q_3=0  $ and $ r_h=0 \Rightarrow m=0 $ (dotdashed), EVH case with  $ q_2=q_3=1  $ and $ r_h=0.5 \Rightarrow q_1=0.16 $ and $ m=3.27 $ (solid line), and extremal 2-charge case with $ q_1=0, q_2=q_3=1  $ and $ r_h=0 \Rightarrow m=m_c=1 $ (dashed) are shown.}
\end{figure}
\begin{equation}\label{63}
\alpha-\frac{\sum_{i=1}^3 \kappa_i(\omega+j^i)^2}{f^2}=0,~~~~~~~~~~\Pi^2=\frac{\kappa_1}{f^2}I_1=\frac{\kappa_1^2(\omega+j^1)^2}{f^2}.
\end{equation}
Finally, one can gain the drag force as
\begin{equation}\label{64}
F_{\psi}=-\frac{1}{2\pi \alpha'}\Pi'=-\frac{1}{2\pi \alpha'}\frac{\kappa_1(\omega+j^1)}{f}.
\end{equation}
The point associated with the near horizon behavior of string tail can be interesting for both non-extremal and extremal cases. In non-extremal cases, $ \alpha $ and $ \beta $ can be expanded as
\begin{equation}
\alpha (r) =\alpha '(r_h)(r-r_h)+\mathcal{O}(r-r_h)^2,~~~~~~~~~~~ \beta (r)=\frac{\beta _{-1}}{(r-r_h)}+\mathcal{O}(1).
\end{equation}
Regarding $ \psi '^2$, for single charge case the near horizon behavior of $ \psi ' $ should be proportional to $ (r-r_h)^{-1} $ so that $ \psi $ would be logarithmically divergent, whereas for 3-charge case we have $ \psi '\propto (r-r_h)^{-1/2} $ and there is no divergency from $ \psi $. For extremal case, the expansion of $ \alpha $ and $ \beta $ is as follows
\begin{equation}
\alpha (r) =\alpha ''(r_h)(r-r_h)^2+\mathcal{O}(r-r_h)^3,~~~~~~~~~~~ \beta (r)=\frac{\beta _{-2}}{(r-r_h)^2}+\mathcal{O}(r-r_h)^{-1}.
\end{equation}
In this case for single charge, $ \psi '$ in the near horizon would be proportionate to $ (r-r_h)^{-2} $. This behavior for 3-charge case would be $ \psi '\propto (r-r_h)^{-1} $ and it has logarithmic divergence.
\subsection{EVH limit}
We are interested in calculating the drag force (\ref{64}) for different extremal black holes with different R-charges (see Fig. \ref{f3} ). We take into account black holes with the limit of Extremal Vanishing Horizon (EVH) \cite{deBoer:2011zt}. To attain the limit of EVH for a black hole, temperature and entropy or related area of horizon should set to zero,  $ T=0 $, $ A_h=0 $, while their ratio, $ A_h/T $, keeps finite. From $ T=0 $, we find
\begin{equation}\label{65}
q_1=\frac{r_h^4L^2+2r_h^6+r_h^4(q_2+q_3)}{m_c L^2-r_h^4}=r_h^4 C
\end{equation}
where $ m_c=q_2q_3/L^2 $, and from $ h(r_h)=0 $ we have
\begin{equation}\label{66}
m-m_c=2r^2_hm_c C+\frac{r_h^4}{L^2}[(q_2+q_3)C-1].
\end{equation}
We can also obtain extremal 2-charge case through two approaches. The first approach complies setting $ q_1 $ and $ r_h $ to zero, taking $ q_2=q_3 $. For the second one, one can attain $ q_1 $ as a function of $ r_h $ and $ q_2 $, for which we can apply the relation obtained in (\ref{65}). Then, by tending $ r_h $ to zero, we would end up in the case. By putting these quantities into (\ref{64}), we would find that these two approaches would arrive at the same result for the drag force. Another case we would like to study is BPS single charge. This one is an extremal black hole gained by $ q_2=q_3=0 $ which this leads to $ m=0 $. We computed the drag force for each of the extremal cases we discussed as a function of $ \omega $, shown in Fig. \ref{f3}.
\section{conclusion}
In this paper we studied drag force which is experience by a moving pointlike probe moving in a strongly coupled gaued theory at zero temparature and finite density. To do so, we used the holography method and considered the behavior of a trailing string in the bulk for extremal geometries. First by looking at the RNAdS background and calculating the drag force for the extremal case, we find the drag force for all velocities without any constraint, as apposed to \cite{Gubser:2009qf}. The IR regime was studied by finding behavior of trailing string in the near horizon limit. One finds in contrast with the non-extremal case, there is no logarithmic divergency for the extremal case. We also take into account the stochastic behavior of the probe in this background. We attain the temperature of the worldsheet for the extremal case and compare it to that of non-extremal one for two different chemical potentials. It turns out this temperature in the extremal type should be always smaller than non-extremal case with $ \mu ^2=1 $ and it also holds for $ \mu ^2=3 $  by velocities around $ 0.55 $. Langevin coefficients are computed and it is found they satisfy the universal bound $ \kappa_{L}/\kappa_{T}>1 $ for this extremal background. This bound was studied in \cite{Giataganas:2013hwa} and it was shown that it is violated for anisotropic backgrounds.

We generalize these calculations to 5-dimensional 3QBH geometry, considering vanishing entropy limit. Getting extremality for 2QBH can be obtained from two different approaches, and it is found that the drag force would be the same in these approaches. Besides, the Langevin coefficients are computed in this extremal case and it is worthwhile to mention that for ultra-relativistic velocities the ratio  $ \kappa_{L}/\kappa_{T} $ becomes less than one and does not obey the universal property. Finally, we investigate drag force in 10 dimensional 3QBH and consider the motion of string in $ S^5 $ part directions, specifically in the $ \psi $ direction. The near horizon behavior of the string is compared between non-extremal and extremal cases. In the non-extremal and 1-charge case it turns out that $ \psi $ has logarithmic divergency while for the 3-charge case there is no divergency. In the extremal and single charge case $ \psi \propto (r-r_h)^{-1} $ and for the 3-charge case, it is shown that there is logarithmic divergency. It is an interesting observation that the IR divergency depends on the charges of the background. We ultimately calculate the drag force and find it numerically for EVH, extremal 2-charge, and BPS single charge cases. It would be interesting to study potential behavior of particle and antiparticle at zero temperature and finite density using extremal geometries.
\section*{Acknowledgment}
The authors would like to acknowledge the scientific atmosphere of school of physics of IPM, Tehran, during the course of this project.


 \end{document}